\documentstyle[prd,aps,floats]{revtex}

\begin{document}
\preprint{astro-ph/9912349}
\draft

%
%
\input epsf

\renewcommand{\topfraction}{0.99}
\renewcommand{\bottomfraction}{0.99}

\twocolumn[\hsize\textwidth\columnwidth\hsize\csname
@twocolumnfalse\endcsname

\title{Perturbation amplitude in isocurvature inflation scenarios}
\author{Andrew R.~Liddle and Anupam Mazumdar}
\address{Astrophysics Group, The Blackett Laboratory, Imperial College, London 
SW7 2BZ, United Kingdom} 
\date{\today} 
\maketitle
\begin{abstract}
We make a detailed calculation of the amplitude of isocurvature perturbations 
arising from inflationary models in which the cold dark matter is represented by 
a scalar field which acquires perturbations during inflation. We use this to 
compute the 
normalization to large-angle microwave background anisotropies. Unlike the case 
of adiabatic perturbations, the normalization to COBE fixes the spectral 
index of the perturbations; if adiabatic perturbations are negligible then 
$n_{{\rm iso}} \simeq 0.4$. Such blue spectra are also favoured by other 
observational data. Although the pure isocurvature models are unlikely to 
adequately fit the entire observational data set, these results also have 
implications for models with mixed adiabatic and isocurvature perturbations.
\end{abstract}

\pacs{PACS numbers: 98.80.Cq, 98.70.Vc \hspace*{6.1cm} astro-ph/9912349}

\vskip2pc]

\section{Introduction}

There has been recent interest \cite{lm,Peeb,peeb2,pi,ss,langlois} in 
inflationary models 
which give rise to 
isocurvature perturbations in the Universe, rather than the traditional 
adiabatic ones. The simplest way to achieve this is to assume that the material 
presently making up the cold dark matter takes the form of a massive scalar 
field, which during inflation received perturbations by the usual mechanism. 
With suitable subsequent evolution, these perturbations can be the dominant ones 
in the present Universe.

The familiar inflationary calculation for adiabatic perturbations is extremely
simple (see Ref.~\cite{LL} for a review), because once the perturbations have
been stretched to wavelengths greater than the Hubble length they become
constant, and no calculation is required until the much later epoch when they
cross back inside the Hubble radius, which for scales of interest is well after
nucleosynthesis.  By contrast, isocurvature perturbations \cite{lk,eb,kofman} 
are 
free to
evolve on superhorizon scales, and the amplitude at the present day depends on
the details of the entire cosmological evolution from the time that they are
formed.  On the other hand, because all super-Hubble radius perturbations evolve
in the same way, the {\em shape} of the isocurvature perturbation spectrum is
preserved during this evolution and so need not be calculated.  In the
literature, therefore, comparison with observation \cite{peeb2,pi} 
has largely focussed on the
spectral index of the perturbations, and it has been assumed that the model
parameters could always be tweaked to ensure the amplitude is correct.
However, as we will see the constraints on amplitude and slope are intimately 
related in these scenarios. In this paper, we compute the Sachs--Wolfe 
contribution to large-angle microwave 
background anisotropies for a particular class of isocurvature models which give 
nongaussian perturbations. Unlike the case of adiabatic perturbations, the 
Sachs--Wolfe effect fixes the slope of the isocurvature perturbations, as well 
as their amplitude, because the slope determines the matter density.

\section{Model description}

The class of models we discuss is characterized by the cold dark matter being a 
scalar field with a time-dependent mass, which apart from its quantum 
fluctuations sits in the vacuum state during the entire history of the Universe. 
Particular implementations of that general statement have been given by Linde 
and Mukhanov \cite{lm} and by Peebles \cite{Peeb}. It is well known that a 
sufficiently massive scalar 
field $\phi$, with $m \gg H$ where $H$ is the Hubble parameter, oscillates with 
decaying amplitude such that its energy density $\rho_\phi \propto 1/a^3$ where 
$a$ is the scale factor, and further that perturbations in this field obey the 
same equations as perturbations in a pressureless fluid. Such a scalar field is 
therefore a viable cold dark matter candidate, the classic example being the 
axion (see Ref.~\cite{kofman} for an early treatment and Ref.~\cite{KT} for a 
review).
 
The mass of the scalar field in such a scenario must have quite a complicated 
evolution in order to generate a viable model.

\subsection{During inflation} 

If the field is effectively massless, $m \ll H$, its perturbations will be 
scale-invariant. Such a spectrum is excluded by comparison of COBE data with 
galaxy clustering; what is needed is a spectrum with a tilt favouring 
short-scale perturbations (usually called a blue spectrum). This can 
successfully be generated if $m \simeq H$ at the relevant stages of inflation 
\cite{lm}; 
the mass causes perturbations to die away outside the horizon, and those which 
have been outside the longest receive the greatest suppression. However if $m 
\gg H$ during inflation, then perturbations will be highly suppressed with an 
$n_{{\rm iso}} = 3$ spectrum,\footnote{We use conventions where $n_{{\rm iso}} = 
0$ corresponds to scale-invariance. Unfortunately the literature features at 
least three different conventions, where scale-invariance can correspond to an 
index of $-3$, $0$ or $1$.} which is also unsatisfactory. One must therefore 
tune the mass to be close to $H$, perhaps best done by generating an 
effective mass $m^2 _{{\rm eff}} = \alpha H^2$ where $\alpha$ is some coupling 
hoped to be of order one. 

\subsection{After inflation} 

We are seeking to create isocurvature perturbations, of the type where the CDM 
perturbation is cancelled by the collective perturbation of all other species. 
In fact, the perturbations we are discussing are actually isothermal, since the 
fluctuations in $\phi$ are uncorrelated with those in the inflaton field which 
is what ultimately determines the radiation density after reheating. These 
isothermal perturbations can be broken into an adiabatic and an isocurvature 
part. However, as long as the energy density of the CDM field is totally 
subdominant to that of the inflaton at the end of inflation, the isocurvature 
perturbation will be by far the dominant part \cite{Peeb}. 

Another important point is that the 
inflaton must not decay into the CDM field, which would destroy the isocurvature 
perturbations, and so must decay only into conventional particles. Therefore at 
the end of inflation the CDM field has much less energy density than everything 
else, whereas by the present it has to dominate the baryonic component (e.g.~the 
present Universe may have $\Omega_{{\rm baryon}} \simeq 0.04$, $\Omega_\phi 
\simeq 0.3$ and a cosmological constant). While the field has a significant mass 
then 
$\rho_\phi \propto 1/a^3$, as compared to the thermalized fluid which at early 
times goes as $\rho_{{\rm rad}} \propto 1/a^4$. However it turns out that a more 
dramatic 
gain is required, which is achieved by having an epoch where the mass is small 
enough to be negligible. When this happens, the scalar field feels no force and 
is effectively fixed, leading to a constant $\rho_\phi$ enabling it to `catch 
up' with the conventional matter.

\subsection{At late times}

In order to preserve the successes of the standard cosmology, the field 
preferably should already be acting as cold dark matter by the time of 
nucleosynthesis (though there may be no observational conflicts even if this 
behaviour is delayed until sometime later). This requires 
that the mass is greater than the Hubble parameter at that time, and this 
guarantees that it will act this way when any astrophysically-interesting 
perturbations enter the horizon. 
Depending on 
whether one believes in a cosmological constant or not, the desired dark matter 
density is between about 5 and 20 times the baryon density, so the field must 
start to scale as $1/a^3$ just when it reaches that relative density. This 
requires fine-tuning, but that's true of all dark matter models as they are all 
trying to explain the same coincidence of dark matter and baryon densities. The 
simplest scenario is that the mass is the bare mass of the field; the effective 
mass was large during inflation due to some coupling, and then the field evolved 
as effectively massless until $H$ fell to be of order this bare mass.

An interesting feature of this model is that the perturbations are crucial to 
define the background. In normal models there is a homogeneous background 
density 
$\rho_{{\rm CDM}}$ about which the perturbations are defined, but here the 
homogeneous background is $\phi = 0 \Rightarrow \rho_\phi = 0$. Instead, the 
mean density of the dark matter is generated by the perturbations, taking 
advantage of the density being proportional to the square of the field 
perturbation. At each point in space the field $\phi$ is oscillating with some 
amplitude. If we define $\Phi({\bf x})$ to be the maximum of the oscillation at 
point ${\bf x}$, then the energy density is 
\begin{equation}
\rho_\phi({\bf x}) = \frac{1}{2} m^2 \Phi^2({\bf x}) \,.
\end{equation}
and the mean density is
\begin{eqnarray}
\label{ener}
\bar{\rho}_\phi = \frac{1}{2} m^2 \langle \Phi^2({\bf x}) \rangle & =
	& m^2 \langle \phi^2({\bf x}) \rangle \nonumber \\
 & = &  m^2 \int_0^{k_{{\rm max}}} {\cal P}_{\phi}(k) \, 
	\frac{dk}{k} \,.
\end{eqnarray}
In that last expression, ${\cal P}$ is the power spectrum of $\phi$, defined as
\begin{equation}
{\cal P}_\phi(k) = \frac{k^3}{2\pi^2} \langle \phi_{{\bf k}}^* \phi_{{\bf k}} 
\rangle \,,
\end{equation}
where $\phi_{{\bf k}}$ are the Fourier components of $\phi({\bf x})$ and $k = 
|{\bf k}|$. We have included an upper cut-off $k_{{\rm max}}$, intended to be 
the shortest scale on which there are perturbations and estimated as the Hubble 
scale at the end of inflation.

The background dark matter density therefore has to be obtained 
self-consistently out of the perturbation calculation. Were the spectrum flat, 
this cannot be achieved; we need the power spectrum to rise sharply to short 
scales, as then the correct 
density can be obtained by suitable tuning of $k_{{\rm max}}$ and $m$ to give 
the observed value. Calculationally this is also desirable, as in considering 
the large-scale perturbations one can imagine smoothing out the small-scale 
perturbations to obtain a genuinely smooth background. It is interesting, though 
certainly not compelling, that the blue spectrum is favoured both in order to 
obtain the correct dark matter density and to match large-scale structure 
observations.

One should also note that the statistics of the perturbations are neither 
gaussian nor strictly speaking chi-squared, though they are related to the 
latter. It is the total density $\rho_\phi$ which is chi-squared distributed, 
not the density contrast $\delta({\bf x}) \equiv [\rho_\phi({\bf x}) - 
\bar{\rho}_\phi]/ \bar{\rho}_\phi$.

\section{Background and perturbation evolution}

The complete model needed to evaluate the amplitude of perturbations therefore 
has two inputs. The first is the evolution of the Hubble parameter $H(t)$. 
During inflation this is governed by the inflaton field, which is some other 
scalar field whose energy is dominating over the CDM field. After inflation, it 
is determined by whatever the matter content of the Universe is, and we shall be 
assuming the standard, simplest evolution of radiation domination after 
inflation giving way to matter domination. One could include a cosmological 
constant too but we will ignore that minor complication. The second input is the 
time-dependent mass of the CDM field. As discussed, this will be of order $H$ 
during the late stages of inflation, then at the end of inflation drops to a 
small constant value which is initially negligible. Then eventually the Hubble 
parameter drops below this value and the field begins truly to behave as CDM.

The perturbations need to be tracked at various stages. First the amplitude 
arising from quantum fluctuations needs to be computed. Subsequently, the 
perturbation is outside the Hubble radius, but still decays as it has a 
significant mass $m \sim H$. At the end of inflation the mass becomes negligible 
and the field freezes out, before eventually the mass becomes important and the 
field begins to evolve again. For all the earlier stages, it is simplest to 
define the evolution in terms of the scalar field itself. However, at the last 
stage we have exactly a standard isocurvature CDM model, and standard results 
(for example for the Sachs--Wolfe effect) can be applied. These are usually 
expressed not in terms of the field, but rather in terms of the entropy 
perturbation $S$ defined later on.

To begin with we need to study perturbations in the CDM field 
which is responsible for generating isocurvature fluctuations.
The background evolution is governed by the inflaton field, which dictates that
the Hubble expansion during inflation is $H^2 \simeq V/3M_{\rm Pl}^2$,
where $\rm V$ is the potential for the inflaton field and $M_{{\rm Pl}} = 
\sqrt{1/8\pi G}$ is the reduced Planck mass. As explained earlier, we
assume that energy density of the CDM field $\phi$ is subdominant
compared to that of the inflaton during the slow-roll phase. We 
are interested in studying the fluctuations generated in the $\phi $ field.
The fluctuations in the inflaton
field cause perturbations in the curvature giving rise to   
adiabatic perturbations, but such perturbations are to be arranged to be 
negligible compared to the isocurvature ones. 

There are various sources which would produce an effective mass to the CDM
field \cite{lm}.  One possibility is the existence of a time-varying coupling of 
the CDM
to the inflaton field.  During inflation such a coupling can be adjusted to give
an effective mass to the CDM which is proportional to the Hubble parameter.
Another possibility is the existence of a non-minimal coupling of the CDM to
gravity.  Such a term appears as a correction to the mass of the CDM and does
not alter the dynamics of the overall evolution as the CDM
field is oscillating at the bottom of the potential $ \phi =0$.  Such a
coupling gives a correction to the mass of the CDM field:  $m^2(H)= m^2 + 12 \xi
H^2$ \cite{lm}.  For our purposes, it does not matter what the source of the
correction is as long as it fixes the effective mass to be of the order of the
Hubble parameter.  During inflation the presence of such a term provides the
tilt in the power spectrum and just after the end of inflation it must switch
off to avoid the CDM field dominating too early.

To study the perturbations we need the perturbation equation
for the CDM
\begin{eqnarray}
\label{pert}
\ddot{\delta \phi}_{k} +3H \dot{\delta \phi}_{k} + \left[\frac{k^2}{a^2}+
m^2 _{\rm {eff}}(t)\right]\delta \phi_{k} =0\,,
\end{eqnarray}
where the effective mass $m_{\rm {eff}}(t)$ is
\begin{eqnarray}
\label{eff}
m^2_{\rm {eff}}(t) = m^2 + \alpha (t) H^2\,,
\end{eqnarray}
where $m$ is the bare mass term of the CDM field. We assume that
$m^2 \ll \alpha (t)H^2$ during inflation. The solution of Eq.~(\ref{pert}) for 
$k \ll aH$ is well 
known \cite{mfb}
\begin{eqnarray}
\label{sol0}
\delta \phi_{\rm k} \approx H \left(Ha \right )^{-3/2}\left[\frac{k}{Ha}
\right]^{-\sqrt{9/4-m^2 _{\rm {eff}}/H^2}}\,.
\end{eqnarray}
If the ratio of effective mass $m_{\rm {eff}}(t)$ and $H$ is constant 
during the entire 
period of inflation, and assuming $H$ is suitably slowly varying, then at 
the end of inflation the solution for the power spectrum for the modes which are 
already outside the horizon is \cite{lm}
\begin{eqnarray}
\label{sol}
{\cal P}_{\delta \phi}(k) \equiv \frac{k^3}{2\pi^2}
|\delta \phi_{k}|^2 \approx \frac{1}{2\pi^2}
H_{\rm e}^2 \left( \frac{k}{H_{\rm e}a_{\rm e}}\right )^{2\alpha/3}\,,
\end{eqnarray}
where the subscript `e' denotes the end of inflation.

This final 
expression, Eq.~(\ref{sol}), is strictly
correct only at the end of inflation, and we have replaced $m_{\rm {eff}}$ with 
$\alpha H^2$, where $\alpha$ is a constant less than one. An alternative 
physical scenario would be if $m_{\rm{eff}}$ were constant rather than $\alpha$, 
in which case the tilt will pick up an extra scale-dependence, depending on the 
details of the explicit model. We will only consider the situation when $\alpha$ 
is a constant. Even in that case, an extra scale-dependence of the tilt arises 
because of variation of $H$ during inflation, but this contribution will be 
insignificant compared to the tilt induced by the mass.

After inflation ends, the effective mass must drop precipitously from its value 
during inflation to the bare mass $m$ which is then much less than $H$. 
Initially then the mass can be neglected, and Eq.~(\ref{pert})
has a simple solution on large scales: $\delta \phi_{k} ={\rm constant}$. As a 
result 
the power spectrum, Eq.~(\ref{sol}), remains unchanged.

This simple situation does not hold for ever, as
the Hubble parameter drops as $a^{-2}$ during radiation era, where $a$ is
the scale factor of the Universe. Eventually the Hubble parameter becomes 
equal to the bare mass of $\phi$ and henceforth the mass of the
$\phi$ field cannot be neglected in Eq.~(\ref{pert}). From this time 
onwards, the CDM field acts like a true massive CDM field, and the amplitude of 
the perturbations
falls as $a^{-3/2}$ during all subsequent phases of the Universe.
Noting that 
$H(t) = H_{\rm e} a_{{\rm e}}^2/a^2$ during radiation domination,  we find that 
at any time 
after mass turn on the spectrum of the fluctuations is given by
\begin{equation}
\label{evol}
{\cal P}_{\delta \phi}(k)  
\approx \frac{1}{2\pi^2} \left(\frac{a}{a_{\rm e}}\right)^{-3}
\left(\frac{H_{\rm {e}}}{m}\right)^{3/2} H_{\rm e}^2 
\left( \frac{k}{H_{\rm e}a_{\rm e}}
\right )^{2\alpha/3}\,.
\end{equation}
Preferably this transition should take place before the synthesis of light
elements, when the ratio between the photon and the baryon is 
already a fixed constant, though compatibility with observations may still be 
maintained even if it happens some time afterwards. This expression for the 
spectrum continues to hold at all later epochs, 
including matter domination.

After mass turn-on, it is useful to consider an alternative quantity, the 
entropy perturbation $S$ given by 
\begin{equation}
\label{ent0}
S = \frac{\delta \rho_{\phi}}{\rho_{\phi}} - \frac{3}{4}\frac{\delta 
\rho_{{\rm r}}}{\rho_{{\rm r}}}\,,
\end{equation}
where the subscript `r' denotes radiation. Because the perturbations are 
actually isothermal (albeit in a limit where isothermal and isocurvature are 
almost the same thing), we have $\delta \rho_{{\rm r}} = 0$ and this can be 
written as
\begin{eqnarray}
\label{ent1}
S = \frac{\delta \rho_{\phi}}{\rho_{\phi}} \equiv 
\frac{\rho_{\phi}({x}) - \bar \rho_{\phi}}{\bar \rho_{\phi}} =
\frac{\delta \phi ^{2}({x}) - \langle \delta \phi ^{2}({x})\rangle }
{\langle \delta \phi^2({x})\rangle }\,,
\end{eqnarray}
where $\bar \rho_{\phi}$ is the mean energy density and it is defined as  
the spatial average of the field fluctuations. $\bar \rho_{\phi} (x)$ defines
the total energy density of the field $\phi $ at any given time and it 
is determined using Eq.~(\ref{ener}) from the field fluctuations themselves. The 
advantage of using the entropy perturbation is that it is a standard result 
\cite{mfb,LL} that on super-horizon scales it continues to be conserved even 
when the radiation is no longer dominant (at which point the radiation will 
acquire perturbations from the influence of the CDM field fluctuations on the 
expansion rate), and this conservation law can be used to evolve the 
perturbations up to decoupling to compute the Sachs--Wolfe effect.

Now we have all the necessary formulae to determine the Sachs--Wolfe effect for 
the isocurvature fluctuations of the $\phi $ field.

\section{Comparison with observations}

\subsection{Normalization from the Sachs--Wolfe effect}

The anisotropy due to the entropy perturbation is given by the well-known 
formula \cite{eb,star,LL}   
\begin{equation}
\label{con}
\frac{\Delta T(\bf{x})}{T} = - \frac{2 S}{5} 
\end{equation}
Following the same steps as for the adiabatic case (see e.g.~Ref.~\cite{LL}), 
the spectrum of the cmb, denoted by $C_\ell$, is given by  
\begin{equation}
\label{c2}
C_\ell = \frac{16\pi}{25} \int_{0}^{\infty} \frac{dk}{k} {\cal P}_{S}(k) 
j^2_\ell
\left( \frac{2k}{H_{0}a_0} \right)\,,
\end{equation}
where ${\cal P}_{S}(k) \delta({\bf k}-{\bf k}^{\prime}) \equiv 
\langle S_{{\bf k}}^* S_{{\bf k}^{\prime}} \rangle k^3/2\pi^2$, and $j_\ell$ is 
a spherical 
Bessel function. We have taken the distance to the last-scattering surface as 
$2/a_0 H_0$, where the subscript $0$ corresponds to the
present value. This formula is strictly true only for 
a critical-density universe, but could be readily generalized to low-density. 

From Eq.~(\ref{ent1}), except for the ${\bf k} = 0$ mode we have
\begin{equation}
S_{{\bf k}} = \frac{1}{\langle  \delta\phi^2 \rangle} (\delta\phi^2)_{{\bf k}} 
\,, \quad ({\bf k} \neq 0) \,.
\end{equation}
Hence
\begin{equation}
{\cal P}_S(k) = \frac{1}{\langle \delta \phi^2 \rangle^2} {\cal 
P}_{\delta\phi^2}(k) = 
\frac{m^4}{\bar{\rho}_\phi^2}  {\cal P}_{\delta\phi^2}(k) \,.
\end{equation}

To evaluate $C_\ell$ we need to compute the power spectrum of $\delta \phi^2$. 
Its calculation is quite subtle, and we carry it out in the Appendix. The 
result, for the range of tilts we are interested in, is  
\begin{equation}
\label{int1}
{\cal P}_{\delta \phi^2} = \frac{I(2\alpha/3)}{4 
\pi^4}\left(\frac{a_e}{a}\right)^{6}
\left( \frac{H_e}{m}\right )^3 \frac{{H_e}^4}{1-2\alpha /3} \left(\frac{k}
{H_e a_e}\right)^{4\alpha /3} \,,
\end{equation}
where $I(2\alpha/3)$ is a constant described in the Appendix, which is of order 
10 for 
the values of $\alpha$ we will be interested in.
Here $H_{\rm e}$ is the Hubble parameter at the end of 
inflation.  

Substituting Eq.~(\ref{int1}) into Eq.~(\ref{c2}), we get the following 
expression
\begin{eqnarray}
\label{cl}
C_\ell& =&
\frac{4 \, I(2\alpha/3) \, m^4}{25 \pi^3 \bar \rho^2 _{\phi}}
\left(\frac{a}{a_{\rm e}}\right)^{-6}
\left(\frac{H_{\rm e}}{m}\right)^{3}\frac{H_{\rm e}^4}{1-2\alpha /3} \times 
\nonumber \\
&&\frac{1}{(H_{\rm e}
a_{\rm e})^{4\alpha/3}}  
\int_{0}^{\infty}dk k^{4\alpha/3} j^2_\ell \left(\frac{2k}{H_{0}a_0}
\right) \,,
\end{eqnarray}
Eq.~(\ref{cl}) can be further simplified by performing the integration
\begin{eqnarray}
\label{cl1}
\ell (\ell +1)C_\ell & = & \frac{2 \, I(2\alpha/3) \, m^4 H^4_{\rm e}}{25 
\pi^3\bar \rho^2 
_{\phi}(1-2\alpha /3)}
\, \frac{a_{\rm e}^6}{a^6} \,
\frac{H_{\rm e}^3}{m^3}\,
\left(\frac{H_{0}a_0}{2H_{\rm e}a_{\rm e}}\right)^{4\alpha/3} \times
\nonumber \\
&& \hspace*{-1.8cm}  \left\{ \frac{\sqrt{\pi}}{2} \ell(\ell +1)
\frac{\Gamma[(2-4\alpha /3)/2]
\Gamma[\ell +2\alpha /3]}{\Gamma[(3-4\alpha/3)/2] \Gamma[\ell+(2-2\alpha /3)]}
\right\}.
\end{eqnarray}
In fact, as long as $\ell \gg 1$ and $\ell \gg 2\alpha/3$, the complicated term 
within the braces can be approximated simply by $\ell^{4\alpha/3}$ (see 
e.g.~Ref.~\cite{LL}).

Although in this expression the quantities $\bar{\rho}_\phi$ and $a$ were given 
at the time of decoupling, since $\bar{\rho}_\phi \propto 1/a^3$ they can in 
fact be evaluated at any epoch, including the present.

Next we can estimate $\bar \rho_{\phi}$ by using Eqs.~(\ref{ener}) and 
(\ref{evol}), 
obtaining
\begin{equation}
\label{ener1}
\bar \rho_{\phi} =\frac{3m^2}{4\pi^2 \alpha} \left(\frac{a}{a_{\rm e}}\right)
^{-3}
\left(\frac{H_{\rm {e}}}{m}\right)^{3/2} H_{\rm e}^2
\left( \frac{k_{\rm {max}}}{H_{\rm e}a_{\rm e}}
\right )^{2\alpha/3}\,.
\end{equation}
Substituting $\bar \rho_{\phi}$ in Eq.~(\ref{cl1}), and taking $k_{\rm{max}} 
\sim 
H_{\rm e}a_{\rm e}$ since we assume that the shortest scale perturbations 
generated are at the horizon scale at the end of inflation, one gets a simple 
relation for a 
multipole of order $\ell$, namely
\begin{eqnarray}
\label{cl2}
\ell (\ell +1)C_{\ell} \approx \frac{32 \pi}{255} \, \frac{I(2\alpha/3) \,
\alpha^2}{1-2\alpha /3} \left[
\frac{\ell H_{0}a_{0}}{2H_{\rm e}a_{\rm e}}\right]^{4\alpha/3}\,.
\end{eqnarray}

The COBE observations \cite{ben} give the $10$th multipole as 
$\ell(\ell +1)C_{\ell} \sim 7\times 10^{-10}$. Taking the ratio of Hubble scales 
at the present and at the end of inflation to be given by the usual estimate
$\ln\left(H_{\rm e}a_{\rm e}/H_0a_0\right) \approx 50$, this gives
\begin{equation}
n_{{\rm iso}} \simeq \frac{4\alpha}{3} \simeq 0.43 \,, 
\end{equation}
where $n_{{\rm iso}}$ refers to the slope of ${\cal P}_S$.

Predominantly, then, the Sachs--Wolfe effect for isocurvature perturbations 
fixes the {\em slope} of the perturbations, rather than the amplitude as one 
finds for adiabatic perturbations. The reason is that the slope, favouring 
short wavelengths, is what determines the matter density, and for the observed 
perturbations $S \simeq 10^{-5}$ to be achieved, the tilt must be steep 
enough to favour short-scale perturbations in determining the matter density 
against which the perturbations are observed. Were the spectrum flatter, then 
the background density would not be high enough, whereas if it were steeper than 
the large-scale perturbations would not give enough density contrast against the 
mean background derived from the short scales. This argument was given in 
Ref.~\cite{lm}; our calculation had given a precise quantification of the 
constraint.

It is intriguing that the slope required to give the correct magnitude of the 
Sachs--Wolfe effect is also in better agreement as to the slope of the 
large-angle $C_\ell$ than a flat spectrum, although it seems unlikely that a 
pure isocurvature model will be able to fit the complete set of observational 
data \cite{SGB,pi}.

\subsection{The CDM matter density}

We can constrain the bare mass $m$ of the CDM from the present density
\begin{eqnarray}
\label{cric}
\bar{\rho}_\phi \simeq \rho_{{\rm crit}} = 
{3H_{0}^2 M_{\rm Pl}^2}\,,
\end{eqnarray}
where we continue to assume that we live in a 
critical-density universe. With the help of Eq.~(\ref{ener1}) 
we can derive a simple expression for the CDM mass
\begin{equation}
\label{mass}
m = (4 \pi^2 \alpha )^2 M_{\rm Pl}^4 \frac{H_0^4}{H_{\rm e}^7}
\left( \frac{a_0}{a_{{\rm e}}}\right)^{6}\,, 
\end{equation}
where we took $k_{\rm {max}} = H_{\rm e} a_{\rm e}$.To simplify this, we model 
the Universe as matter-dominated during reheating until $H_{{\rm rh}}$, followed 
by the usual cosmology. During matter domination $a \propto H^{-2/3}$ and during 
radiation domination $a \propto H^{-1/2}$, leading to 
\begin{eqnarray}
m & = & 16 \pi^4 \alpha^2 M_{\rm Pl}^4\frac{H_0^4}{H_{\rm e}^7}
\left( \frac{a_0}{a_{{\rm eq}}}\, \frac{a_{{\rm eq}}}{a_{\rm {rh}}} \,
\frac{a_{\rm{rh}}}{a_{\rm e}}\right)^{6} \nonumber \\
& = & 16 \pi^4 \alpha^2 M_{\rm Pl}^4 \,\frac{H_{{\rm eq}}}{H_{{\rm rh}} \, 
H_{{\rm e}}^3}
 \,.
\end{eqnarray}
For critical density the Hubble parameter at matter--radiation equality is 
$H_{{\rm eq}} = 5.25 \times 10^6 \, h^3 H_0$ where $H_0 = 2.1 h \times 10^{-42}$ 
GeV (with $0.5 < h < 0.75$) is the present Hubble parameter. Putting in 
characteristic values leads to
\begin{equation} 
\label{mass2}
\frac{m}{1 \, {\rm GeV}} = 44 \alpha^2 \, g_*^{-1/2} \, h^4 \, 
\left(\frac{10^{13} \, {\rm GeV}}{H_{{\rm e}}}\right)^3 \, \left( \frac{10^{10} 
\,{\rm GeV}}{T_{{\rm rh}}}\right)^2 \,.
\end{equation}
where $g_* \sim 100$ is the number of particle species at reheating.

The model appears compatible with observations for a reasonable range of 
parameters. Firstly, gravitational wave production requires $H_{e} < 10^{14}$ 
GeV, or gravitational waves would give the dominant contribution to the COBE 
anisotropies (see e.g.~\cite{LL}). Indeed, it will need to be some way less than 
this to make sure that adiabatic perturbations do not exceed the isocurvature 
ones which we presume to dominate. The mass $m$ needs to be less than $H_{{\rm 
e}}$ for the scenario to be consistent, and should not turn on too late if the 
standard cosmology is to be recovered. Everything would be completely safe if 
turn on precedes nucleosynthesis, but may continue to be permitted even if turn 
on is later, since the dark matter density would be negligible during 
nucleosynthesis. However since $H_{{\rm nucl}} \simeq 10^{-24} \, {\rm GeV}$ 
this is unlikely to be the situation. Finally, if ones assumes supersymmetry 
then the 
reheat temperature is constrained by gravitino production from two-body 
scattering processes in the thermal bath. If the reheat temperature is too high 
then gravitinos are produced with an abundance which destroys standard 
nucleosynthesis. To avoid this, it should satisfy be $T_{\rm rh} \leq 10^{10} \, 
{\rm GeV}$ \cite{sa}. This implies reheating should be inefficient, though 
the problem is no different here to usual inflation models, and may indeed be 
less severe than some standard models as we can accommodate a low energy density 
at the end of inflation.

Parameters exist which satisfy all these constraints. For example, $H_{{\rm e}} 
\simeq 10^{12} \, {\rm GeV}$, $T_{{\rm rh}} \simeq 10^9 \, {\rm GeV}$ and $m 
\sim 10^4\, {\rm GeV}$ is a suitable combination.

\section{Discussion and conclusions}

We have carried out a full calculation of the amplitude of microwave 
anisotropies within a 
class of isocurvature inflation models which includes models by Linde and 
Mukhanov \cite{lm} and be Peebles \cite{Peeb}. We have not dwelt on the 
model-building issues of obtaining the appropriate time-dependence of the CDM 
field's mass, but under the assumption that it can be made to behave as required 
we have evaluated the model constraints. The key result is that the Sachs--Wolfe 
effect in these models primarily constrains the slope, rather than the 
amplitude, of the perturbations. Typically observational comparisons have 
assumed that the slope is a free parameter (e.g.~\cite{pi}), without taking this 
additional restriction into account.

A broad range of parameters appears capable of simultaneously matching the COBE 
observations and the required CDM matter density. It is intriguing that the 
slope $n_{{\rm iso}} \simeq 0.4$ required to match the Sachs--Wolfe amplitude is 
also in the right vicinity to have a chance of matching the slope of the 
large-angle anisotropy spectrum. Nevertheless, it is probably true that such 
pure isocurvature scenarios are unable to fit the complete data set 
\cite{SGB,pi}. There does however remain scope for mixed adiabatic and 
isocurvature scenarios \cite{SGB,langlois,mixed,pi}, and our results can be 
applied, with 
minor modification, in those scenarios too.

\section*{Acknowledgments}
A.M. is supported by Inlaks foundation. We thank Antonio da Silva, Lev Kofman, 
David Lyth 
and David Wands for useful discussions. 

\appendix
\section*{The power spectrum of $\delta\phi^2$}

The entropy perturbation is given by the square of the field perturbation, and 
so we need the spectrum of that quantity. We assume that the power spectrum of 
$\delta\phi$ itself is given by
\begin{equation}
\label{specdefa}
{\cal P}_{\delta\phi}(k) = A \, \left(\frac{k}{H_{{\rm e}} a_{{\rm e}}} 
\right)^n \,.
\end{equation}
The spectrum of $\delta \phi^2$ is given by \cite{lyth}\footnote{In 
Ref.~\cite{lyth} an unusual mix of Fourier conventions is used, but the final 
result is correct.}
\begin{eqnarray}
\label{int}
{\cal P}_{\delta \phi ^2}(k) = \frac{k^3}{2\pi}\int_{0}^{a_{{\rm e}} H_{{\rm 
e}}} \frac{{\cal P}_{\delta \phi}(|{\bf p}|){\cal P}_{\delta \phi}(|{\bf k}-{\bf 
p}|)}{|{\bf p}|^3 |{\bf k}-{\bf p}|^3} d^3 {\bf p}\,,
\end{eqnarray}
where $|{\bf p}|$ and $|{\bf k}-{\bf p}|$ are the wavenumbers. This integral 
needs to be carried out with some care concerning the angular part. We can use 
rotational symmetry to align ${\bf k}$ along the 
$z$-axis to make $|{\bf k} - {\bf p}|$ independent of the azimuthal angle, so 
that $d^3{\bf p} \equiv 2\pi p^2 \sin \theta \, d\theta \, dp$. 
Substituting in the spectrum from Eq.~(\ref{specdefa}) gives
\begin{equation}
{\cal P}_{\delta \phi ^2} = \frac{k^3 A^2}{(a_{{\rm e}} H_{{\rm e}})^{2n}}
\int_0^{a_{{\rm e}} H_{{\rm e}}} \int_0^\pi p^{n-1} |{\bf k}-{\bf p}|^{n-3} \sin 
\theta \, d\theta \, dp \,,
\end{equation}
where $p \equiv |{\bf p}|$ and $k = |{\bf k}|$. Using
\begin{equation}
|{\bf k} - {\bf p}|^2 = |{\bf k}|^2 + |{\bf p}|^2 - 2 |{\bf k}|\,|{\bf p}|
	\cos \theta \,,
\end{equation}
leads to the integral 
\begin{eqnarray}
{\cal P}_{\delta \phi ^2}(k) & = & \frac{k^2 A^2}{(a_{{\rm e}} H_{{\rm 
e}})^{2n}} \frac{1}{1-n} \times \\
& & \int_0^{a_{{\rm e}} H_{{\rm e}}} p^{n-2} \left[|k-p|^{n-1} - (k+p)^{n-1} 
\right] \, dp  \,. \nonumber
\end{eqnarray}
Finally, setting $u = p/k$ gives 
\begin{eqnarray}
{\cal P}_{\delta \phi ^2}(k) & = & \frac{A^2}{1-n} \left(\frac{k}{a_{{\rm e}} 
H_{{\rm e}}}\right)^{2n} \times \\
&& \int_0^{a_{{\rm e}} H_{{\rm e}}/k} u^{n-2} \left[ |1-u|^{n-1} - (1+u)^{n-1} 
\right] du \,. \nonumber
\end{eqnarray} 
where we are interested in $n$ positive, and usually less than one.
This final integral has interesting properties. It is convergent as $u \sim 0$. 
For $n<1$ the integrand is divergent at $u = 1$ (i.e.~$p=k$) but it is 
integrable. Finally, at large $u$ the integrand goes as $u^{2n-4}$. If $n 
\gtrsim 3/2$, the upper limit dominates and one finds ${\cal P}_{\delta \phi ^2} 
\propto k^3$; this limit appears for instance in the analysis of preheating 
models in Ref.~\cite{LLMW} where $n=3$. However for the smaller $n$ values we 
are 
interested in, the integral is dominated by $u \sim 1$. The upper limit of the 
integral becomes irrelevant and can be taken to infinity for modes with $k 
\lesssim a_{{\rm e}} H_{{\rm e}}$, so that the integral is independent of $k$. 
The spectrum therefore has tilt ${\cal P}_{\delta \phi ^2}(k) \propto k^{2n}$. 
For the range of $n$ we will be interested in, $0 < 2n < 1$, we can write
\begin{equation}
{\cal P}_{\delta \phi ^2}(k) = \frac{A^2 I(n) }{1-n} \left(\frac{k}{a_{{\rm e}} 
H_{{\rm e}}}\right)^{2n} \,,
\end{equation}
where
\begin{equation}
\label{integral}
I(n) = \int_0^\infty u^{n-2} \left[ |1-u|^{n-1} - (1+u)^{n-1} 
\right] du \,.
\end{equation}
We are not aware of an analytical evaluation of this integral except for special 
cases, but it is readily done numerically and we show the result in Fig.~1. 

\begin{figure}[t]
\centering 
\leavevmode\epsfysize=6cm \epsfbox{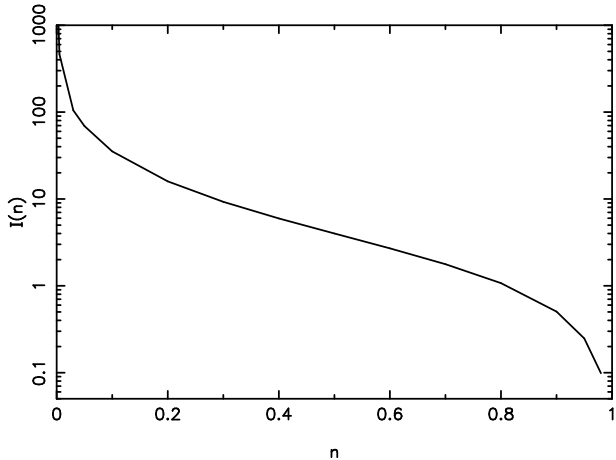}\\
\caption[integral]{\label{f:integral} The result of numerical evaluation of the 
integral of Eq.~(\ref{integral}), for $n$ in the range zero to one.}
\end{figure}

 
\end{document}